\documentclass[]{aastex631}

\let\tablenum\relax	
\usepackage{bm}
\usepackage{siunitx}
\usepackage{ulem}

\newcommand{\msun}{M_\odot}

\newcommand{\fiducial}{$28$}
\newcommand{\ranozo}{$27$}
\newcommand{\ranoten}{$40$}
\newcommand{\rafbone}{$18$}
\newcommand{\rafbzo}{$17$}
\newcommand{\rafbten}{$31$}
\newcommand{\denoone}{$22$}
\newcommand{\denozo}{$15$}
\newcommand{\denoten}{$46$}
\newcommand{\defbone}{$1.1$}
\newcommand{\defbzo}{$1.5$}
\newcommand{\defbten}{$9.4$}

\newcommand{\fiduciallong}{$28$}
\newcommand{\ranozolong}{$26$}
\newcommand{\ranotenlong}{$33$}
\newcommand{\rafbonelong}{$18$}
\newcommand{\rafbzolong}{$17$}
\newcommand{\rafbtenlong}{$24$}
\newcommand{\denoonelong}{$20$}
\newcommand{\denozolong}{$14$}
\newcommand{\denotenlong}{$36$}
\newcommand{\defbonelong}{$0.85$}
\newcommand{\defbzolong}{$0.82$}
\newcommand{\defbtenlong}{$2.4$}

\newcommand{\mindet}{\defbone}
\newcommand{\maxdet}{\denoten}
\received{December 9, 2021}
\revised{February 4, 2022}
\accepted{February 7, 2022}

\shorttitle{Detectability of BHBs with Gaia}
\shortauthors{Shikauchi et al.}
\graphicspath{{./}{./}}

\begin{document}

\title{Detectablity of Black Hole Binaries with {\it Gaia}: Dependence on Binary Evolution Models}

\correspondingauthor{Minori Shikauchi}
\email{shikauchi@resceu.s.u-tokyo.ac.jp}

\author{Minori Shikauchi}
\affiliation{Department of Physics, the University of Tokyo, 
7-3-1 Hongo, Bunkyo, Tokyo 113-0033, Japan}
\affiliation{Research Center for the Early Universe (RESCEU), the University of Tokyo,
7-3-1 Hongo, Bunkyo, Tokyo 113-0033, Japan}

\author{Ataru Tanikawa}
\affiliation{Department of Earth Science and Astronomy, College of Arts and Sciences, the University of Tokyo, 
3-8-1 Komaba, Meguro, Tokyo 153-8902, Japan}

\author{Norita Kawanaka}
\affiliation{Department of Astronomy, Graduate School of Science, Kyoto University, Kitashirakawa Oiwake-cho, Sakyo-ku, Kyoto, 606-8502, Japan}
\affiliation{Hakubi Center, Kyoto University, Yoshida-honmachi, Sakyo-ku, Kyoto, 606-8501, Japan}
\affiliation{Yukawa Institute for Theoretical Physics, Kyoto University, Kitashirakawa Oiwake-cho, Sakyo-ku, Kyoto, 606-8502, Japan}



\begin{abstract}
Astrometric satellite {\it Gaia} is expected to observe
non-interacting black hole (BH) binaries with luminous companions
(LCs) (hereafter BH-LC binaries), a different population from BH X-ray
binaries previously discovered.  The detectability of BH-LC binaries
with {\it Gaia} might be dependent on binary evolution models.
We investigated the {\it Gaia}'s detectability of BH-LC binaries 
formed through isolated binary evolution 
by means of binary population synthesis technique, and
examined its dependence on single and binary star models: supernova models, common envelope (CE) ejection
efficiency $\alpha$, and BH natal kick models.  We estimated that \mindet~-- \maxdet~BH-LC binaries
can be detected within five-year observation, and found that $\alpha$ has
the largest impacts on the detectable number.  In each model,
observable and intrinsic BH-LC binaries have similar distributions.
Therefore, we found three important implications:
(1) if the lower BH mass gap is not intrinsic (\textit{i.e.} $3$ -- $5$ $\msun$ BHs exist), {\it{Gaia}} will observe $\leq 5 \msun$ BHs, 
(2) we may observe short orbital period binaries with light LCs if CE efficiency is significantly 
high, 
and (3) we may be able to identify the existence of natal kick from eccentricity distribution.
\end{abstract}

\keywords{astrometry --- stars: black holes --- binaries: general}


\section{Introduction}
Stellar mass black holes (BHs) are formed at the end of massive stars.
Some previous researches estimated that there are $10^8$ -- $10^9$ stellar mass BHs in the Milky Way (MW) \citep{Shapiro1983,vandenHeuvel1992,BrownandBethe1994,Samland1998,Agoletal2002}.
Some of them have been detected as X-ray binaries in the MW and they have very short periods such as several hours. 
Less than 100 \citep{BlackCAT2016} BHs were detected in this way, much less than theoretically predicted.

Gravitational wave (GW) observation is an alternative way to find binaries including BHs.
In extragalactic distances, more than 60 binary BHs/ neutron star(NS)-BH binaries have been detected \citep{GWTC2019, GWTC-3}. 
Thus, BH searches so far discovered BHs in short period binaries, which have electromagnetic and/or GW emissions through interactions between BHs and their companions. 

There are two other ways to search for binaries consisting of BHs and luminous companions (LCs) (hereafter, BH-LC binaries) with longer orbital periods: radial velocity observations and astrometric observations. There are already a few detection reports of long orbital period BH-LC binaries with radial velocity search \citep{Giesers2018,Thompson2019,Liu2019}. 
In particular, \cite{Liu2019} have reported a $70 \msun$ BH in a binary system \citep[but see][]{ Eldridgeetal2020,Tanikawaetal2020,Safarzadeh2019,El-Badry2020,Irrgang2020}. 
\cite{Thompson2019} found a non-interacting giant star-unseen object binary by combining radial velocity and photometric data. 
Since the mass of unseen object is estimated to be $3.3_{-0.7}^{+2.8}$ $\msun$, it should be a light BH or massive NS, although the result is under debate \citep{vandenHeuvelandTauris2020,Thompsonetal2020}. 
Also, by employing radial velocity and photometric data, \cite{Jayasingheetal2021} reported a binary consisting of a red giant V723 Mon and a BH candidate with its mass of $3.04 \pm 0.06 \msun$.
In astrometric observations, \cite{GouldandSalim2002} investigated BHs not producing supernovae (SNe) for the first time 
by using data sets of Hipparcos and indicated that the successor will observe their companion stars.
As the successor of {\it Hipparcos}, {\it Gaia} mission \citep{GaiaColl2016} was launched in 2013 and has been providing 
information about parallaxes and proper motions of $2000$ million stars with higher precision such as $\mu$as \footnote{https://www.cosmos.esa.int/web/gaia/science-performance}. 
Now, {\it Gaia} Early Data Release 3 (EDR3) has been released \footnote{https://www.cosmos.esa.int/web/gaia/earlydr3}, and the next data release (DR3) including full information of binaries, 
is planned in 2022 \footnote{\url{https://www.cosmos.esa.int/web/gaia/release}}. 
Since the cadence of the observation is about 50 days and {\it Gaia} has been in science mode for several years, 
DR3 may include BH-LC binaries with longer periods than those that can be detected in X-rays or GWs, whose typical periods are hours to years.
Moreover, it can observe a dark compact objects such as a NS, a white dwarf and a brown dwarf and LC binaries with a good precision \citep{Andrewsetal2019}.

Some previous researches have estimated hundreds to thousands of BH-LC binaries can be detected with {\it Gaia} \citep{MashianLoeb2017, Breivik2017, Yamaguchi2018, Yalinewich2018, Kinugawa2018, Shao2019, Chawlaetal2021}. 
All of the binaries so far were thought to be isolated binaries, which
were born as tight binaries and did not experience any dynamical
interactions with other stars in the MW disk.  In stellar clusters,
BH-LC binaries can be formed by dynamical interactions and
\cite{Shikauchietal2020} estimated about 10 BH-main sequence (MS) star binaries can be detected. 
The detectability is greatly affected by binary evolution models and observational constraints they employed.
However, the previous researches investigated how the detectability will be affected by one or two of the binary evolution models: SN models, common envelope (CE) efficiency $\alpha$, and natal kick models. 
In this work, we employ binary population synthesis to see how all the combinations of the binary evolution models affect the detectability, and discuss whether we can give constraints on binary evolution models from current/future observations with stringent observational constraints employed in \cite{Yamaguchi2018}.

The structure of this paper is as follows.
In section \ref{sec:method}, we introduce the binary evolution models used in this work, and we summarize the observational constraints. 
We show the results of binary population synthesis in
section \ref{sec:result} and discuss the difference among our results and previous works in section \ref{sec:discussion}.

\label{sec:intro}

\section{Method}
\label{sec:method}
In this section, we first summarize a initial setup for the simulation:
the spatial distribution of BH-LC binaries and 
initial conditions of binaries in subsection \ref{sec:IniCon}.
We also introduce a binary population synthesis code
and modifications that we added to it in subsection \ref{sec:BSE}.  
Then We summarize how to obtain the number of detectable BH-LC binaries with
{\it Gaia} in subsection \ref{sec:NumEst}. It is based
on \cite{Yamaguchi2018}.  

\subsection{Initial Condition}
\label{sec:IniCon}
On the assumption that we get the same BH-LC population everywhere in the MW disk, the number of binaries at a position, $\bm{x}$, can be expressed as a function of their BH masses, $m_{\rm BH}$, LC masses, $m_{\rm LC}$, orbital periods, $P$, and eccentricities $e$,  
\begin{equation}
    N(m_{\mathrm{BH}},m_{\mathrm{LC}}, P, e, \bm{x}) = 
    \tilde{N}(m_{\mathrm{BH}},m_{\mathrm{LC}}, P, e) \times t_{\mathrm{LC}} \times \dot{\rho}(\bm{x}),
\end{equation}
where $\tilde{N}(m_{\mathrm{BH}},m_{\mathrm{LC}}, P)$ is 
\begin{equation}
    \tilde{N}(m_{\mathrm{BH}},m_{\mathrm{LC}}, P, e) = 
	N_{\mathrm{sim}}(m_{\mathrm{BH}},m_{\mathrm{LC}}, P, e) / M_{\mathrm{ini,int}},
\end{equation}
where $N_{\mathrm{sim}}(m_{\mathrm{BH}},m_{\mathrm{LC}}, P, e)$ is the
number of binaries with $m_{\mathrm{BH}}$, $m_{\mathrm{LC}}$, $P$, and $e$
in our simulations, $t_{\mathrm{LC}}$ is the lifetime of LC,
$\dot{\rho}(\bm{x})$ is a formation rate density of stars in the MW
($M_\odot$pc$^{-3}$ year$^{-1}$), and $M_{\mathrm{ini,int}}$ is the intrinsic total mass
of binaries in one realization.  
Note that the value $M_{\mathrm{ini, int}}$ is not equal to the total initial mass in our simulation, $M_{\mathrm{ini,sim}}$: we simulate the evolution of binaries whose minimum primary mass is $8 \msun$, while the actual minimum primary mass should be as small as $0.08 \msun$, so the initial total mass of binaries in our simulation is smaller than that in reality.
Considering that $M_\mathrm{ini,sim}$ is $3.1 \times 10^7 \msun$, we adopt $M_\mathrm{ini,int}=1.5 \times 10^8 \msun$.

In this work, we consider only the MW disk because the
interstellar extinction makes it difficult to observe the binaries in
the bulge.  We assume that a local star-formation rate density is
proportional to a local stellar density, and that star formation rate,
$R_{\rm SF}$, is constant everywhere in the MW disk. We adopt a total
star formation rate in the entire MW disk to $R_{\mathrm{SF}}$,
$3.5 \msun\; $\SI{}{ year^{-1}} \citep{OShaughnessyetal2006b}.  Then,
a formation rate density $\dot{\rho}(\bm{x})$ is given by
\begin{equation}
    \dot{\rho}(\bm{x}) = R_{\mathrm{SF}} \times n_{\mathrm{MW}}(\bm{x}),
\end{equation}
where $n_{\mathrm{MW}}(\bm{x})$ is a stellar number density distribution at a position $\bm{x}$ in the MW. 
In the MW disk, we assume that stars are exponentially distributed from the center of the galaxy \citep{deVaucouleurs1959,Kormendy1977a} 
and perpendicular to the galaxy plane \citep{BahcallSoneira1980}. 
The stellar number density distribution $n_{\mathrm{MW}}(\bm{x})$ 
can then be expressed as
\begin{equation}
    n_{\mathrm{MW}}(\bm{x}(r, z)) = n_0 \exp \left ( -{{r - r_0}\over{r_h}} - {{z}\over{h_z}} \right ),
\end{equation}
where $r_0=8.5~{\rm kpc}$ is the distance from the center of the Galaxy to the sun, and $r_h=3.5~{\rm kpc}$ and $h_z=250~{\rm pc}$ are the scale lengths along the $r-$ and $z-$ directions, respectively. 
We adopt the normalization factor, $n_0$, to satisfy the following equation, 
\begin{equation}
    4 \pi \int^{r_{\mathrm{max}}}_0 rdr \int^{z_{\mathrm{max}}}_0 dz n_{\mathrm{MW}}(\bm{x}) = 1,
\end{equation}
and we set $r_{\mathrm{max}} =$\SI{30}{\kilo pc} and $z_{\mathrm{max}} =$\SI{1}{\kilo pc} \citep[see a review for][]{BlandHawthornGerhard2016}.

\begin{table}[h]
    \centering
    \begin{tabular}{ccc|ccc||ccc|cc} \hline
	    SN model & $\alpha$ & kick & $N_{\rm{det}}$ & $N_{\mathrm{det}}(P > 1 \;\mathrm{year})$ & &	SN model & $\alpha$ & kick & $N_{\rm{det}}$ & $N_\mathrm{det} (P > 1 \;\mathrm{year})$ \\ \hline
		rapid    & $0.1$ & no kick  & \ranozo   & \ranozolong   & & delayed & $0.1$ & no kick & \denozo & \denozolong \\ 
		$\cdots$ & $1.0$ & $\cdots$ & \fiducial & \fiduciallong & & $\cdots$ & $1.0$ & $\cdots$ & \denoone & \denoonelong \\ 
		$\cdots$ & $10$  & $\cdots$ & \ranoten  & \ranotenlong  & & $\cdots$ & $10$ & $\cdots$ & \denoten & \denotenlong \\ 
		$\cdots$ & $0.1$ & FB kick  & \rafbzo   & \rafbzolong   & & $\cdots$ & $0.1$ & FB kick & \defbzo & \defbzolong \\ 
		$\cdots$ & $1.0$ & $\cdots$ & \rafbone  & \rafbonelong  & & $\cdots$ & $1.0$ & $\cdots$ & \defbone & \defbonelong \\ 
		$\cdots$ & $10$  & $\cdots$ & \rafbten  & \rafbtenlong  & & $\cdots$ & $10$  & $\cdots$ & \defbten & \defbtenlong \\ \hline 
	\end{tabular}
	\caption{The detectability of BH-LC binaries with {\it Gaia} with
	different models. The mark ``$\cdots$'' represents a repeat of the
	same model above. We estimated \mindet -- \maxdet~binaries can be
	detected. The value $N_\mathrm{det} (P > 1 \; \mathrm{year})$ indicates the detectability of
	BH-LC binaries with period longer than 1~year.}
	\label{tab:det}
\end{table}

Our initial conditions are following.
We employ Kroupa initial mass function \citep{KroupaIMF2001} as the initial mass function of the primary mass. 
The minimum and maximum masses are set to be $8 \msun$ and $150 \msun$, respectively.
We assume a flat mass ratio distribution from $0$ to $1$ \citep{Kuiper1935,KobulnickyandFryer2007} and obtain the secondary mass.
We set the minimum value of the secondary mass to $0.1 \msun$.
For the distribution of initial semi-major axis, we assume a logarithmically flat distribution from $10 R_{\odot}$ to $10^6 R_{\odot}$.
The initial eccentricity is set to be thermally distributed \citep{Heggie1975}. 
We employ the solar metallicity, $Z = 0.02$ to the metallicity. 

\subsection{Binary Population Synthesis Code and Modifications}
\label{sec:BSE}

In order to simulate the potential population of BH-LC binaries, we
employ binary population synthesis code {\sf BSE} \citep{Hurley2000,Hurley2002}. 
Stellar wind models are updated
to metallicity-dependent ones following \cite{Belczynski2010}.  Using
{\sf BSE}, we simulate the evolution of millions of binaries.  They
start from zero-age MS binary
stars and some physical processes such as mass and angular momentum
transfers, CE phases, BH formations and natal kicks.

Here, we adopt some modifications as follows.  First, we employ
SN models suggested in \cite{Fryeretal2012}.  There are two
models: ``the rapid model'' and ``the delayed model''. 
In the former model, only very few compact remnants whose mass is in
the range of $2-4.5M_{\odot}$ are formed.  The resulting gap in the
compact remnant mass function matches the
observations \citep{Ozeletal2010,Farretal2011}.  On the other hand,
the latter model does not produce such a mass gap.  Indeed, whether
the observed BH mass gap is intrinsic or not is still controversial.
Therefore, we employ these two models to see how they affect both the
detectability and binary parameter distributions such as BH and LC
masses. Second, we employ different CE efficiencies $\alpha$. When the
mass transfer is dynamically unstable during Roche Lobe overflow
(RLOF), a binary will enter CE phase. We use an energy conservation
equation derived in \cite{Webbink1984} to treat CE evolution. One of
the CE parameters $\lambda$ includes an effect of the mass
distribution within the envelope and a contribution from the internal
energy \citep{deCool1990,DewiandTauris2000}. We employ the results
in \cite{Claeysetal2014} to obtain $\lambda$. Also, the other parameter
$\alpha$ is CE efficiency with which the orbital energy is used to
eject the mass donor's envelope. Since the value of $\alpha$ is of great uncertainty \citep{Fragosetal2019,Podsiadlowskietal2003,Kieletal2006,YungelsonandLasota2008,MapelliandGiacobbo2018,Chuetal2021, BroekgaardenandBerger2021}, we adopt a variety of CE efficiencies, $\alpha = 1$, $\alpha = 0.1$, and $\alpha = 10$. 
Finally, for BH natal kick, we consider two cases: no kick and ``fallback (FB) kick''. For FB kick, we reduce a NS natal kick by a factor of $(1 -
f_{\mathrm{fb}})$, where $f_{\mathrm{fb}}$ is a fraction of mass of
fallback matter to all the ejected mass during SN explosion.  NS kick
follows Maxwellian distribution with $\sigma
=$\SI{265}{km. s^{-1}} \citep{Hobbsetal2005}.

In summary, we treat 12 parameter sets with different two SN models, three values of $\alpha$, and two natal kick models. For each parameter set, we simulated $10^6$ binary stars.
In the following sections, we refer each binary evolution model set as ``SN model/the value of CE efficiency/kick model''. For example, ``rapid/$\alpha=1$/no kick model'' indicates the rapid model with CE efficiency $\alpha=1$ and no natal kick.

\subsection{Number Estimation}
\label{sec:NumEst}
In the following, we summarize how we assume the distribution of binaries obtained in {\sf BSE} simulations in the MW (\ref{sec:IniCon}), and the observational constraints for {\it Gaia} (\ref{sec:stellarext}, \ref{sec:obsconst}) following \cite{Yamaguchi2018}. 

Based on the spatial distribution of the binaries in the MW, we calculate the total number of BH-LC binaries detectable with {\it Gaia}, $N_{\rm det}$, by summing up $n(m_{\mathrm{BH}},m_{\mathrm{LC}},P,e,\bm{x})$ over $m_{\mathrm{BH}}$, $m_{\mathrm{LC}}$, $P$, $e$, and a distance to the binary $D$ up to the maximum distance $D_{\mathrm{max}}$,  
\begin{equation}
    N_{\rm det} = \int dm_{\mathrm{BH}} \int dm_{\mathrm{LC}} \int dP 
    \int de \int_{|\bm{x}-\bm{x}_0|<D_{\max}} d^3 \bm{x} N(m_{\mathrm{BH}},m_{\mathrm{LC}},P, e, \bm{x}),
    \label{number_estimation}
\end{equation}
where $\bm{x}_0$ is the position of the Sun.

Note that $D_{\max}$ is dependent on the parameters of the binaries such as $m_{\mathrm{BH}}$, $m_{\mathrm{LC}}$ and orbital separation $a$. 
In the following section, we obtain three constraints on $D_\mathrm{max}$: $D_\mathrm{LC}$ obtained considering interstellar extinction, $D_\Pi$ and $D_a$ for the confident detection of BHs (see below equation \ref{parallax_err} and \ref{sep_err} respectively). 
Therefore, we set $D_\mathrm{max} = \min(D_\mathrm{LC}, D_\Pi, D_a)$. 
In the following sections, we introduce how to evaluate $D_\mathrm{LC}$, $D_\Pi$ and $D_a$. 

\subsubsection{Interstellar Extinction}
\label{sec:stellarext}
Here, we evaluate the maximum distance at which the LC of a binary 
is so luminous that it can be observed with {\it Gaia}, $D_{\rm LC}$. 
We consider interstellar extinction and $D_{\mathrm{LC}}$ is dependent on 
a luminosity of LC, $L_{\mathrm{LC}}$, and an effective temperature, $T_\mathrm{eff,LC}$, satisfying the equation below,
\begin{equation}
    m_{\mathrm{V}}(L_{\mathrm{LC}},T_\mathrm{eff,LC},D_{\mathrm{LC}}) = m_{\mathrm{v,lim}},
\end{equation}
where $m_{\mathrm{v,lim}}$ is the maximum apparent magnitude
with which a binary can be observed with {\it Gaia}. We set $m_{\rm v, lim}=20$ \citep{GaiaColl2016}. 
By following \cite{Yamaguchi2018}, we consider V band instead of the {\it Gaia} band because the color $V-I$ can be less than unity \citep[see Figure 14 of ][]{Jordietal2010}. 
The V-band absolute magnitude of a LC, $M_\mathrm{V}$, can be obtained from $L_{\mathrm{LC}}$ and $T_{\mathrm{eff,LC}}$ taking into account the borometric correction \citep[see equation 1, 10, and Table 1 in][]{Torres2010}.

The apparent magnitude, $m_{\mathrm{V}}$, can be obtained by using the relationship shown below,
\begin{equation}
	M_\mathrm{V}(L_{\mathrm{LC}},T_\mathrm{eff,LC}) = m_\mathrm{v} - 5(2 + \log_{10}D/\mathrm{kpc}) - A_\mathrm{V}(D/\mathrm{kpc}),
\label{ext}
\end{equation}
where $D$ is the distance to the binary and
$A_\mathrm{V}$ is a term of interstellar extinction. 
Considering the average extinction in the MW is $\sim 1$\,mag per \SI{1}{\kilo pc}
in V band \citep{Spitzer1978,Shafter2017}, we assume $A_\mathrm{V} \sim
D/\mathrm{kpc}$. 
Then, we obtain $D_{\mathrm{LC}}$ satisfying the following equation,
\begin{equation}
	M_{\rm V}(L_{\mathrm{LC}},T_\mathrm{eff,LC}) + 5(2 + \log_{10}D_{\rm LC}/\mathrm{kpc}) + D_{\rm LC}/\mathrm{kpc} = m_{\mathrm{V, lim}}.
\end{equation}

\subsubsection{Observational Constraints on Confirmed Detection of BHs}
\label{sec:obsconst}
Through astrometric observations, we can identify a LC with a sinusoidal motion as a part of binary consisting of the visible LC and an unseen object. 
To determine the unseen object as a BH,  
we set the lower limit of the unseen object mass should be larger than
$2M_\odot$, that is, the estimated
mass of the unseen object should satisfy the equation below
\begin{equation}
    m_{\mathrm{BH}} - n \sigma_{\mathrm{BH}} > 2M_{\odot},
\label{BHconst}
\end{equation}
where $m_{\mathrm{BH}}$ is the unseen object mass and
$\sigma_{\mathrm{BH}}$ is its standard error. We set $n=1$
following \cite{Yamaguchi2018}. 
There is a caveat: NSs might be contaminated under the condition we use here.
However, in GW190814 \citep{Abbottetal2020GW190814}, a compact object with $2$ -- $3 \msun$ was detected and it is worth searching for objects within such mass range even if they are not BHs.

The observables of a binary are the LC mass, $m_{\rm LC}$, the orbital period, $P$, 
the angular semi-major axis, $a_*$, and a distance to the binary, $D$. 
They can be expressed in one equation shown below,
\begin{equation}
    {{(m_{\mathrm{LC}} + m_{\mathrm{BH}})^2}\over{m_{\mathrm{BH}}^3}} = {{G}\over{4 \pi^2}} {{P^2}\over{(a_* D)^3}},
\label{binary}
\end{equation}
where $G$ is a gravitational constant. 
Using equation~(\ref{binary}), we relate 
the standard error of BH mass, $\sigma_{\rm BH}$, to standard errors of other binary 
parameters, 
\begin{equation}
     \left( {{\sigma_{\mathrm{BH}}}\over{m_{\mathrm{BH}}}} \right )^2 = \left( {{3}\over{2}} - {{m_{\mathrm{BH}}}\over{m_{\mathrm{BH}}+m_{\mathrm{LC}}}} \right )^{-2} \left[ \left( {{m_{\mathrm{LC}}}\over{m_{\mathrm{BH}}}+m_{\mathrm{LC}}} \right )^2 {{\sigma_{\mathrm{LC}}^2}\over{m_{\mathrm{LC}}^2}} + {{\sigma_{P}^2}\over{P^2}} + {{9}\over{4}} \left( {{\sigma_{a*}^2}\over{a_*^2}} + {{\sigma_{D}^2}\over{D^2}} \right) \right].
     \label{mbh}
\end{equation}
where $\sigma_{\mathrm{LC}}$, $\sigma_{a*}$, $\sigma_P$, and $\sigma_D$ are observational errors of the LC mass, the semi-major axis, the orbital period, and the distance of the binary, respectively.
We assume that observational errors are sufficiently smaller than 
each observable, and we also ignore the correlation among the observables.

In order to confirm a detection, when we assume that each standard error should be smaller than 10\% of the value of each variable, that is, 
\begin{equation}
    {{\sigma_{\mathrm{LC}}}\over{m_{\mathrm{LC}}}} < 0.1, 
    {{\sigma_{P}}\over{P}} < 0.1, {{\sigma_{a*}}\over{a_*}} < 0.1,
    \;\; \mathrm{and } \; \; {{\sigma_{D}}\over{D}} < 0.1,
    \label{const}
\end{equation}
BHs with masses of $\gtrsim 3.4 M_{\odot}$ will be confirmed to be detected. 

\cite{Tetzlaffetal2011} indicated that a typical standard error of 
a stellar mass estimated from its spectral and luminosity 
is smaller than 10 \%. 
Therefore, the first condition in equation~(\ref{const}) can be easily achieved. 
A standard error of an orbital period can be suppressed to $\lesssim 10 \%$ when the observed period is shorter than two-thirds of 
the observational time \citep{ESA1997}. 
Recently \cite{Lucy2014} indicated we may be able to recover binary parameters even when the coverage of orbital period is as low as 40 \% and \cite{ONeiletal2019} proposed a new method to estimate binary parameters when the coverage is less than 40 \%. 
Since {\it Gaia} has been observing for five years, we adopt 10 years to the maximum period. 
For the minimum period, we adopt 50 day, the cadence of {\it Gaia}, to it following \cite{Yamaguchi2018}. 

From the rest conditions, two more constraints can be imposed on $D_{\mathrm{max}}$. 
Since a parallax $\Pi$ is inversely proportional to the distance to a binary, we obtain the following equation,
\begin{equation}
  \frac{\sigma_\Pi}{\Pi} \sim \frac{\sigma_D}{D} < 0.1, 
  \label{eq:ParallaxAndDistance}
\end{equation}
where $\sigma_{\Pi}$ is a standard error of a parallax. 
Based on \cite{GaiaColl2016}, $\sigma_{\Pi}$ in G band can be expressed by the apparent magnitude $m_{\mathrm{v}}$, 
\begin{equation}
    \sigma_\Pi = (-1.631 + 680.8z(m_\mathrm{v}) +
    32.73z(m_{\mathrm{v}})^2)^{1/2} [\mu \mathrm{as}], 
    \label{eq:StandardErrorOfParallax}
\end{equation}
where
\begin{equation}
    z(m_{\mathrm{v}}) = 10^{0.4 (\mathrm{max}[12.09,m_{\mathrm{v}}]-15)}.
\end{equation}
Here, we neglect the dependence on the color $(V-I)$. 
Substituting equation~(\ref{eq:StandardErrorOfParallax}) into
equation~(\ref{eq:ParallaxAndDistance}), we obtain the constraint on 
$D$ as
\begin{equation}
	\left( \frac{D}{\rm kpc} \right) < D_{\Pi} = \frac{10^2}{(-1.631 + 680.8z(m_\mathrm{v}) + 32.73z(m_{\mathrm{v}})^2)^{1/2}}.
  \label{parallax_err}
\end{equation}

With respect to the constraint on a standard error of a semi-major
axis, we obtain the other constraint on $D_{\max}$.  Considering the
semi-major axis of each binary is about the orbital radius on the
celestial sphere, the standard error of semi-major axis,
$\sigma_{a*}$, is $\sim \sigma_{\Pi}$. The constraint derived from the condition of a
semi-major axis can be expressed as
\begin{equation}
        \left( \frac{D}{\rm kpc} \right) < D_{a} = \frac{am_{\rm BH}}{10(m_{\rm BH}+m_{\rm LC}) \sigma_\Pi}.
    \label{sep_err}
\end{equation}

Therefore, by adopting the minimum value among $D_{\mathrm{LC}}, D_{\Pi},$ and $ D_a$ as
$D_{\mathrm{max}}$, we consider all the constraints for confident detection of BHs with {\it Gaia}. We also set the maximum distance as \SI{10}{\kilo pc} without considering any constraints as \cite{Yamaguchi2018} did.

\section{Result}
\label{sec:result}
The detectabilities $N_\mathrm{det}$ with different binary evolution models are shown in
Table \ref{tab:det}. 
We found that \defbone~-- \denoten~BH-LC binaries can be detected.
We also show the detectability of binaries with orbital period longer than $1$~year, 
$N_\mathrm{det}(P > 1\;\mathrm{year})$.
For most of the models, long period binaries are dominant in the detectable ones.

\begin{figure}[h]
	\centering
	\plotone{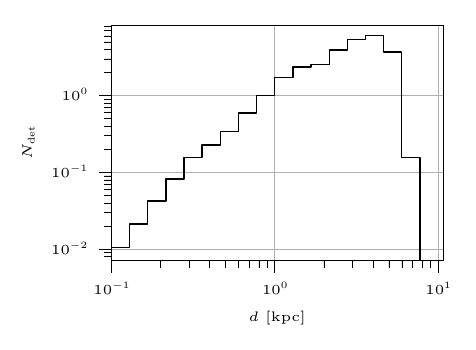}
    \caption{The number of detectable BH-LC binaries $N_\mathrm{det}$ in rapid/$\alpha=1$/no kick model as a function of a distance to them from the Sun, $d$ [kpc].}  \label{kpc_vs_ndet}
\end{figure}

Figure \ref{kpc_vs_ndet} shows the number of detectable BH-LC binaries
$N_\mathrm{det}$ in rapid/$\alpha=1$/no kick model with respect to a
distance to them from the Sun, $d$ [kpc]. The number of detectable
binaries monotonically increases as the integrated volume
does. However, around $d=$ 4~kpc the detectability drastically
decreases because the observational constraints on the orbital
separations are so stringent that farther binaries cannot be detected.

\begin{figure}[h]
	\centering
	\plotone{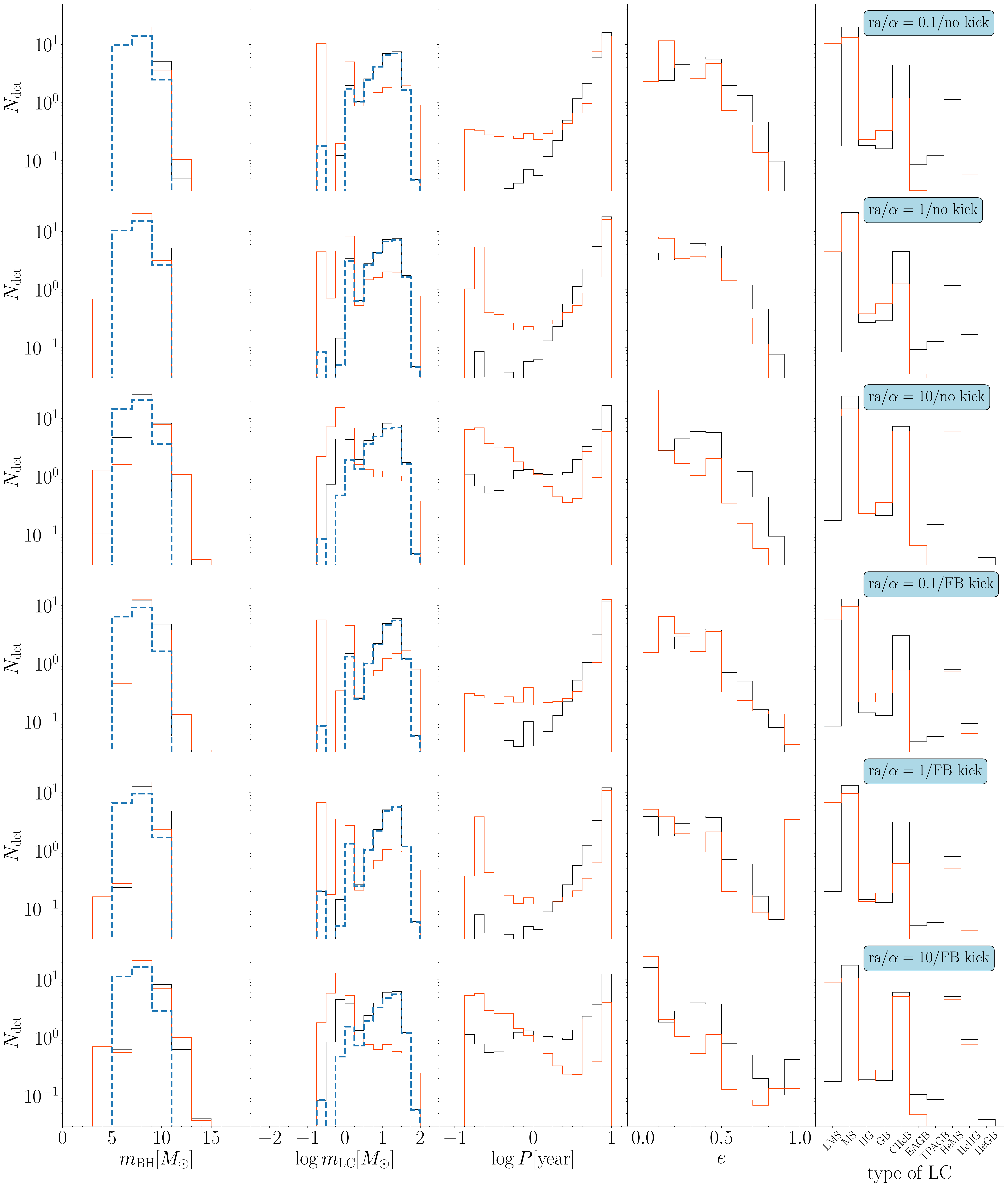}
    \caption{Binary parameter distributions of binaries formed with
	the rapid model. In each row from the top, the distributions
	of $\alpha=0.1$ and no kick, $\alpha=1$ and no kick, $\alpha=10$
	and no kick, $\alpha=0.1$ and FB kick, $\alpha=1$ and FB kick, and
	$\alpha=10$ and FB kick models are shown. Starting on the left,
	there are BH mass, LC mass, orbital period, eccentricity, and LC
	type distributions. In LC type distribution, each abbreviation is following:
	``LMS'': low mass MS star ($m_\mathrm{LC} \lesssim 0.7 \msun$), ``MS'': MS star ($m_\mathrm{LC} \gtrsim 0.7 \msun$), ``HG'': Hertzsprung gap, ``GB'': first giant branch, ``CHeB'': core He burning, ``EAGB'': first asymptotic giant branch, ``TPAGB'': second asymptotic giant branch, ``HeMS'': MS naked He star, ``HeHG'': HG naked He star, and ``HeGB'': GB naked He star. The black line is a
	distribution of detectable binaries and the red one shows the
	Galactic distribution of binaries with $P =$ 50~days to 10
	years normalized in $N_{\mathrm{det}}$.
    In BH mass distribution, the blue dashed line corresponds to a
	distribution obtained from single star evolution. The blue dashed
	line in LC mass distribution shows a distribution of the detectable
	BH-LC binaries with LCs gaining their masses compared to the ZAMS
	stages.}  \label{ra_param}
\end{figure}

\begin{figure}[h]
	\centering
	\plotone{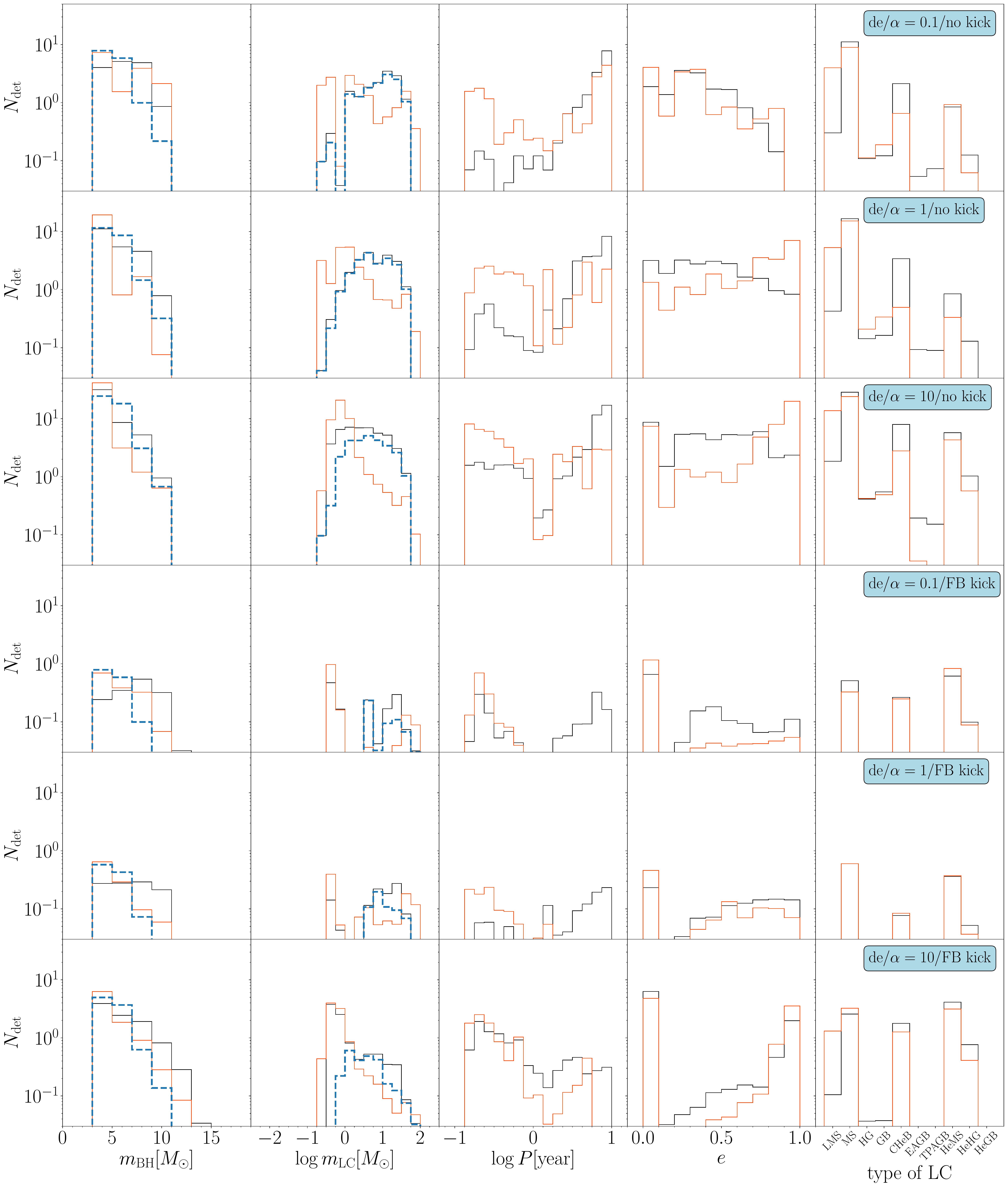}
	\caption{The same as Figure \ref{ra_param} except that SN model is
	the delayed model, not the rapid model.}  \label{de_param}
\end{figure}

\subsection{Predictions of Detectable Binary Parameter Distributions}
\label{sec:param_diff}
Here, we investigate how binary evolution models affect binary parameter distributions 
and discuss whether we can identify them from the observation.

Binary parameter distributions with each binary evolution model are
shown in Figure \ref{ra_param} and \ref{de_param}.
The red line in each panel shows the Galactic distribution of binaries
with $P=50~{\rm days}$ to 10 years, and the black line shows that of
binaries detectable by {\it Gaia}.  Note that for each line the total
number of binaries is normalized to $N_{\rm det}$.  Except for the period
distribution (the third panel from the left), though some small differences exist in each panel,
the detectable binary distributions roughly reflect the features of
the Galactic ones. The period distribution is biased to binaries with long
period such as $P \gtrsim 1$~year compared to the Galactic
distribution. That is because the observational constraint on the
semi-major axis (equation \ref{sep_err}) can be relaxed for looser
binaries.

In the first panel from the left in Figure \ref{ra_param}
and \ref{de_param}, BH mass distribution is shown.  The blue dashed
line depicts a mass distribution of BHs left behind after the death of
a single star. The single stars follow the Kroupa's IMF, and their
evolution is calculated by the single star evolution (SSE)
code \citep{Hurley2000} which uses the same star evolution model as in {\sf BSE}.  The detectable distribution (black) as well
as the Galactic (red) one and one obtained from SSE are very similar.
This can be interpreted in the following way.
In single stellar evolution, BH masses are determined by how much BH progenitors lose their masses through their stellar winds. In binary evolution, BH masses depend on not only strength of stellar winds but also binary interactions. Thus,
BHs formed through binary evolution can have different masses from those through single stellar evolution. However, the effects of stellar winds are much larger than those of binary interactions, since stellar winds are strong due to the solar metallicity. This is why BH mass distributions are similar between the single and binary evolution.

From this result, the BH mass distribution (the first panel from the left) in the rapid model has a peak around 
$8 \msun$, while, in the delayed model, a peak is around $\leq 5 \msun$ and there is no gap structure. 
Therefore, if the lower mass gap is not intrinsic (\textit{i.e.} the intrinsic BH mass distribution is consistent with the delayed model), \textit{Gaia} will observe BHs with mass in that range. We will be able to identify the lower mass gap in BH mass as intrinsic or not from the observation.

In addition, for all the combinations of SN and natal kick models, as CE efficiency becomes high ($\alpha=10$), high mass BHs ($m_\mathrm{BH} \gtrsim 11 \msun$) are formed and become more observable compared to low CE efficiency models. 
The high CE efficiency can make more binaries avoid mergers in CE phases, and evolve to tight (but detectable) BH-LC binaries. In such tight BH-LC binaries, the BHs accrete masses from their companions through mass transfer, and grow to such heavy BHs. Note that the BH growth rate is smaller than the Eddington mass accretion rate.

From LC mass distributions (the second panel from the left), 
the detectable distribution is slightly biased to heavy LCs.
This trend is seen in all the models. 
This is because heavier LCs are more luminous and easier to detect. 
In addition, for both SN models with $\alpha=10$, 
some light LCs ($m_\mathrm{LC} \leq 0.3 \msun$) become detectable 
compared to other CE efficiency models. 
This is because, thanks to the high efficiency, binaries with light LCs expel the envelope efficiently through CE phase and can survive as tight BH-LC binaries. 
Thus, a significant number of short period binaries with light LCs are formed and become detectable as seen in the third panels from left. 
Thus, when we observe short period binaries with light LCs, 
CE efficiency should be significantly large. 

For all the models, the blue lines in LC mass distributions indicate that most of the detectable LCs gain their masses by binary interaction. 
Therefore, the LCs may retain some footprints of the BH progenitors and we expect that detailed observations of LCs can make it clear how the BH-LC binaries were formed. Such footprints were reported in low-mass X-ray binaries (LMXBs) accompanying BHs as chemical anomalies of LC companions \citep[see][for a review]{Casares2017}\citep{Israelian1999,Orosz2001,GonzalezHernandez2004,GonzalezHernandez2005,GonzalezHernandez2006, GonzalezHernandez2011}. 
Although LCs in detectable BH binaries with \textit{Gaia} may be less polluted than those in LMXBs because of their wider separations,
detailed observations of LCs in BH binaries will be helpful to elucidate the origins of chemical anomalies of LCs in LMXBs.

For both kick models with the rapid model and no kick model with the delayed
model, the detectable period distributions (the third panel from the left)
are greatly biased to a large value.  
That is because the motion of LCs with large period
binaries can be easily detected.  However, for the delayed model with FB kick,
a shape of the detectable period distribution varies with a choice of
CE efficiency $\alpha$.  For $\alpha=0.1$ case, two peaks are around
$0.2$~year and $5$~year.  Short period binaries evolve from initially
loose and near-circular orbit binaries such as semi-major axis
$a_\mathrm{ini} \sim 10$ -- $600$~AU and eccentricity
$e_\mathrm{ini} \lesssim 0.5$ with heavy secondaries
($m_\mathrm{sec,ini} \sim 8$ -- $20 \msun$).  Thanks to the
inefficient CE efficiency, a significant number of binaries become
very tight and survive after suffering FB kick, resulting in living
long such as for tens of Myr.  Thus, the first peak is seen in short
period region.  Though long period binaries are more likely to be
disrupted by FB kick than short period ones and thus the number of
long period binaries is small, they are easier to detect and the
second peak can be seen in the long period region.  For higher CE
efficiency models ($\alpha=1$ and $10$), most of the tight binaries
are disrupted because the high CE efficiencies prevent them from
becoming tight enough to survive after suffering from FB kick. Thus, the
short period peak disappears and the detectable distribution is biased
to only longer period in $\alpha=1$ case.  However, a significant
number of tight period binaries can be formed and become observable in
$\alpha=10$ case. They evolve from a different population from $\alpha=0.1$ model, 
that is,
initially tight and eccentric orbit binaries ($a_\mathrm{ini} \sim
0.5$ -- $3$~AU and $e_\mathrm{ini} \gtrsim 0.7$) with light
secondaries ($m_\mathrm{sec,ini} \sim 0.25$ -- $5 \msun$).
The initially high eccentricities can make these binaries enter CE phases, since the binary stars touch with each other at their periapsises.
If CE efficiency is
not high (\textit{i.e.} $\alpha=1$), such binaries cannot expel the donors' envelopes and merge.

In the fourth left panels, we can see that the eccentricities of the detectable BH-LC binaries are slightly biased to a large value except for the delayed model with FB kick. 
The rapid model seems to be robust to FB kick because $f_\mathrm{fb}$ in the rapid model 
tends to be larger than that in the delayed model 
and they do not suffer from FB kick very much. 
When binaries experienced CE phase, the orbits become circularized and tighter, and the resulting BH-LC binaries are difficult to observe.  Hence, the eccentricity distribution of detectable BHs is biased towards larger eccentricities. 
In particular, we found that eccentric long period binaries 
such as $P \sim 10$ years tend to preserve their initial eccentricities. 
However, for the delayed model, FB kick tends to disrupt eccentric binaries 
with long periods and tight binaries with circular orbits are likely to survive. Thus, the Galactic eccentricity distribution is biased to zero and we may be able to distinguish the existence of BH natal kick from the eccentricity distribution.

Finally, LC type distributions (the fifth panels from the left) tell us that 
the majority of detectable binaries have MS star companions except for the delayed model with FB kick. 
A significant number of BH-MS star (LC type 1), Core Helium burning (type 4) 
and MS He star (type 7) binaries have long lifetime such as $\geq 100$~Myr, 
greatly contributing to $N_\mathrm{det}$. 
For the delayed model with FB kick, the majority of such long lifetime binaries are disrupted by FB kick since they have light BHs $m_\mathrm{BH} \lesssim 4 \msun$ and greatly suffer from the kick. Thus, the LC type distribution is different from those of the other models.

According to BH mass distribution, our results also indicate that
lower mass gap BH-LC binaries can be formed even in the rapid models, although the expected detectable number is less than unity. By
mass accretion from secondaries, NSs gain their masses and evolve to
low mass BHs with $m_\mathrm{BH} \sim 3 \msun$. 
We do not get high mass BHs with $\sim 20$ $\msun$ like Cygnus
X-1 \citep{Miller-Jones2021}. This is because we take into account
stronger stellar wind mass loss than inferred by the BH mass in Cygnus
X-1.

\subsection{The Effect of a Choice of $P_{\mathrm{min}}$ and $P_{\mathrm{max}}$ on the Detectability}
\label{sec:Pmin_and_max}

\begin{figure}[h]
	\centering
	\plottwo{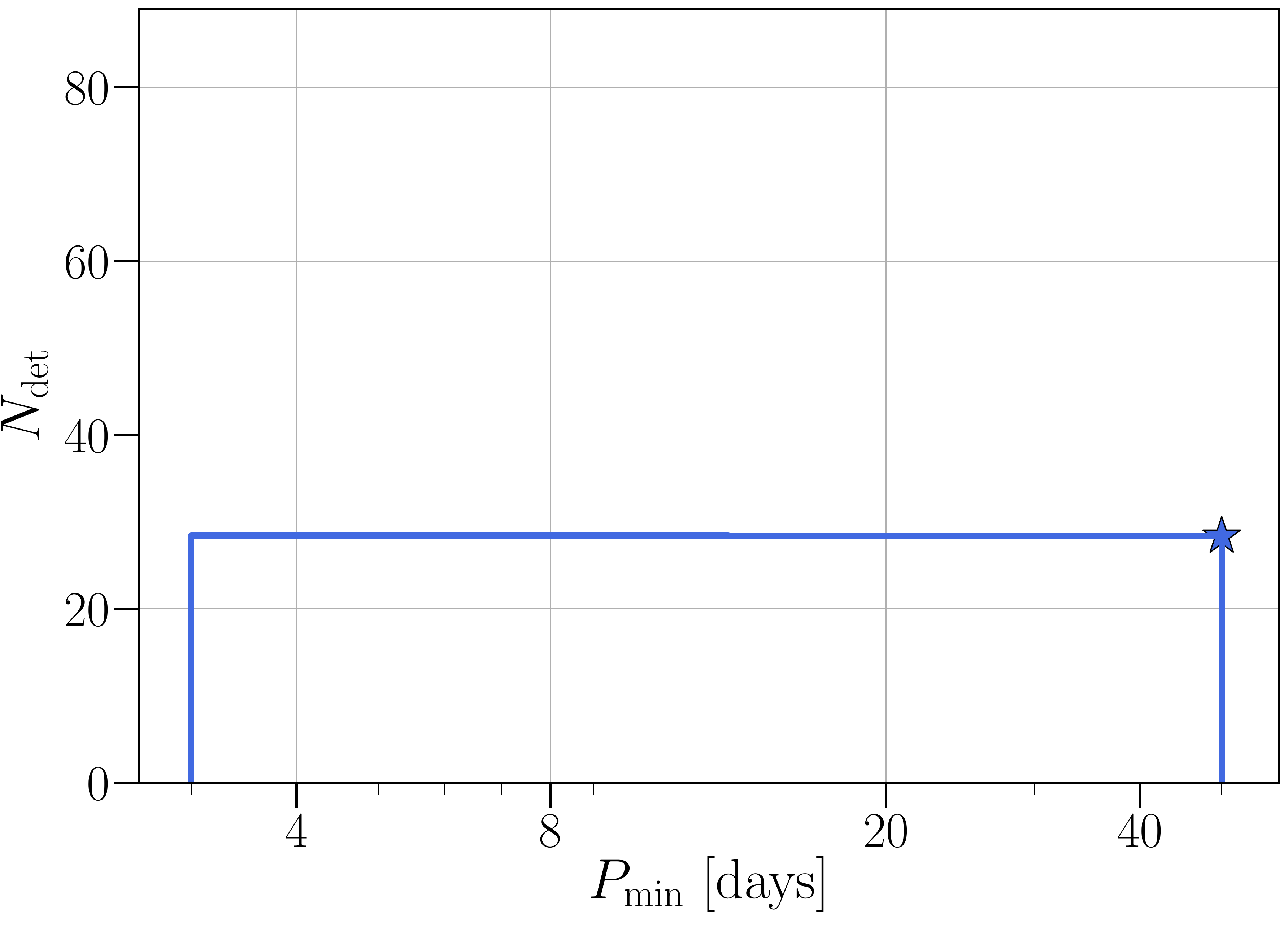}{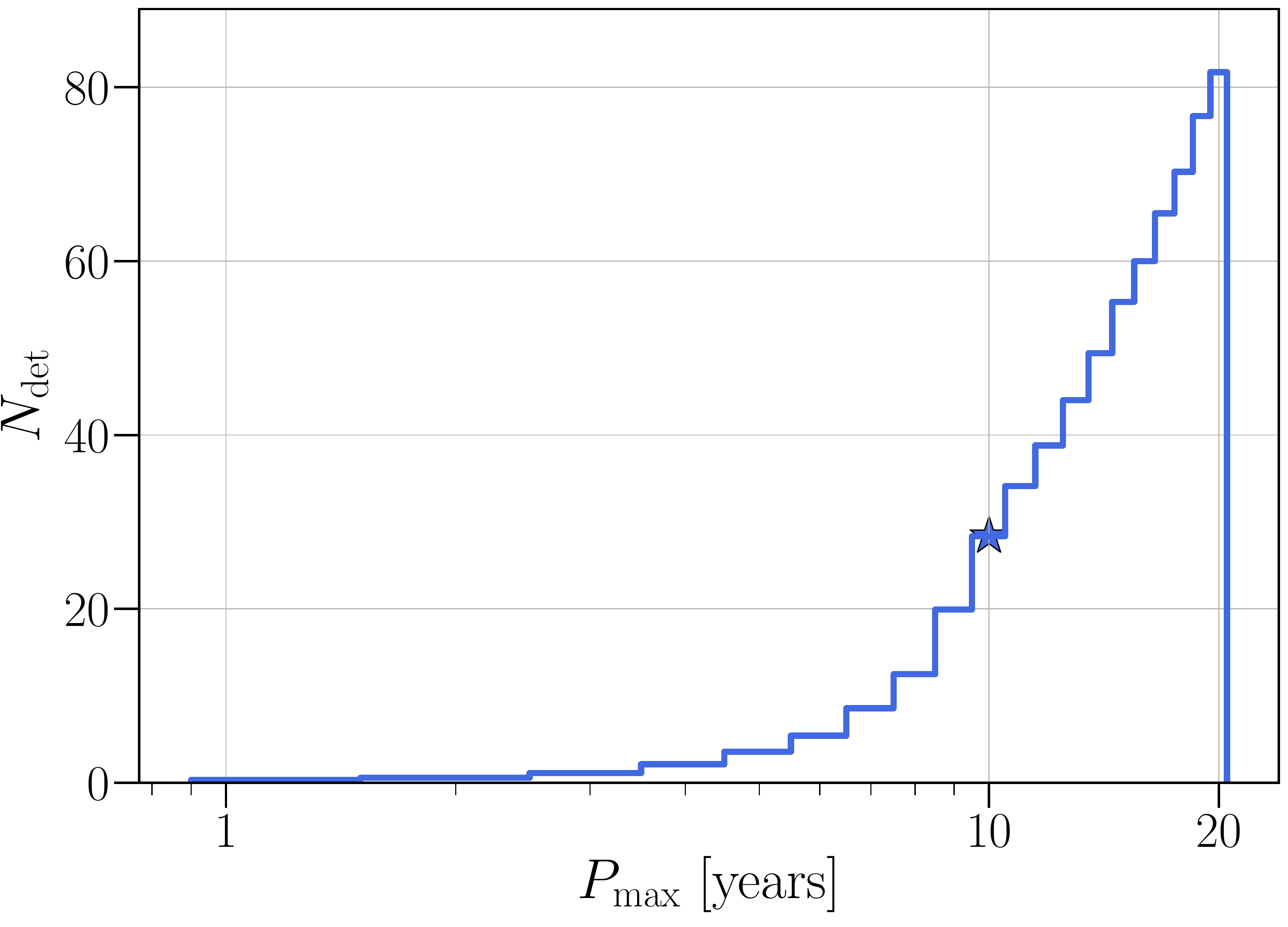}
	\caption{(Left) The relationship between a choice of the minimum
	period $P_\mathrm{min}$ and the detectability $N_{\mathrm{det}}$
	of rapid/$\alpha=1$/no kick model. (Right) The relationship between a choice of
	the maximum period $P_\mathrm{max}$ and the detectability
	$N_{\mathrm{det}}$ of rapid/$\alpha=1$/no kick model. The star marker corresponds to 
	$P_\mathrm{min}$ and $P_\mathrm{max}$ employed so far, e.g. 
	$P_\mathrm{min}=50$~days or $P_\mathrm{max}=10$~years.
	}  \label{n_det_for_P}
\end{figure}

\begin{figure}[h]
	\centering
	\plotone{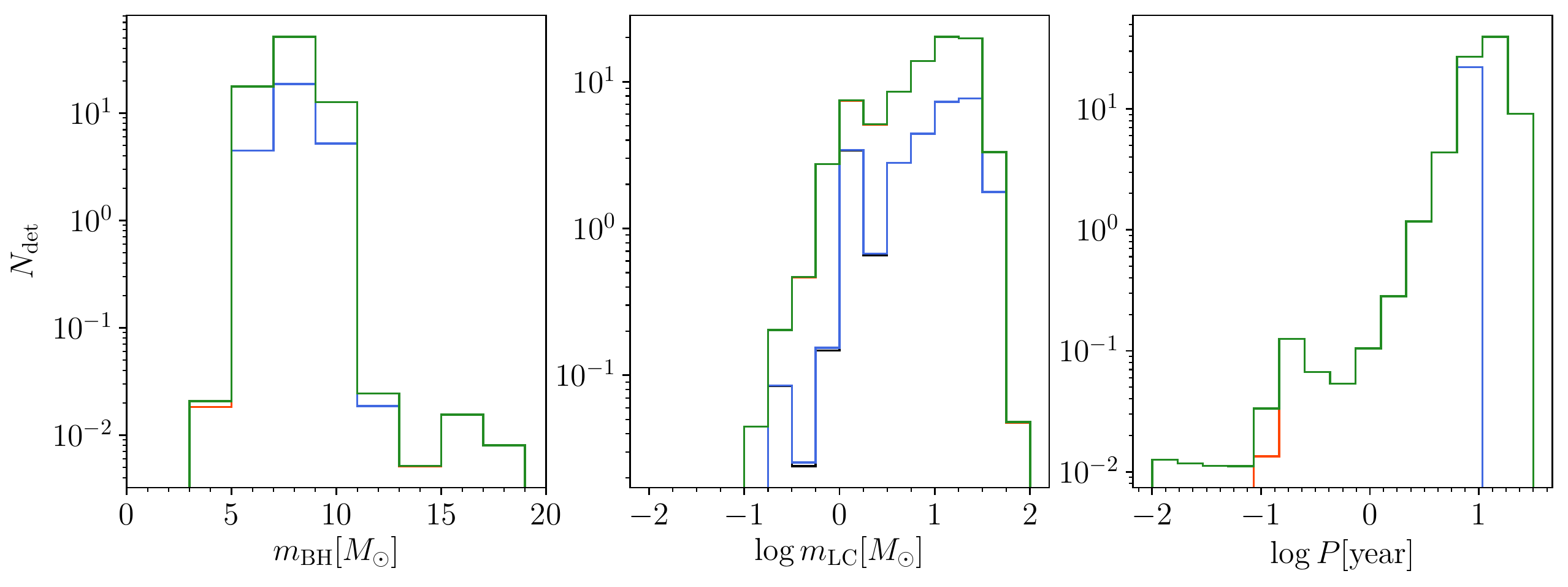}
	\caption{Binary parameter distributions of rapid/$\alpha=1$/no kick model with a various pairs of $P_\mathrm{min}$ and $P_{\mathrm{max}}$, 40~days and 10~years (black), 40~days and 20~years (red), 4~days and 10~years (blue), and 4~days and 20 years (green).}
	\label{param_dist_w_P_min_max}
\end{figure}

In addition to the binary evolution models, we investigate how
choices of $P_\mathrm{min}$ and $P_\mathrm{max}$ affect the
detectability in this section.
BH-LC binaries with rapid/$\alpha=1$/no kick model and a various values of the minimum
period $P_{\mathrm{min}}$, 4, 8, 20, and 40 days. 
The detectability of $P_\mathrm{min}=$4~days increases by at most 0.3 \% compared with that of $P_\mathrm{min}=$40~days. This is \textit{not} because the number of short-period BH-LC binaries is small. Rather, they are difficult to be detected, since their proper motions are small due to their short periods.
The right panel in Figure \ref{n_det_for_P} indicates how the
detectability in rapid/$\alpha=1$/no kick model will change when we
choose a various value of $P_\mathrm{max}$ from 1~year to 20~years.
In the calculations used for Figure \ref{kpc_vs_ndet} to \ref{de_param}, we adopted 10~years as $P_{\mathrm{max}}$. If we extend
it to 20~years, $N_{\mathrm{det}}$ will increase up to $\sim 82$.

In Figure \ref{param_dist_w_P_min_max}, the distributions of BH mass $m_{\mathrm{BH}}$, LC mass $m_{\mathrm{LC}}$ and an orbital period $P$ are shown.
The distributions do not depend on $P_{\mathrm{min}}$ as compared between the black and blue lines, and between the red and green curves. This is because the number of the detectable BH-LC binaries does not increase even if $P_{\mathrm{min}}$ becomes smaller, as described above. Although the number of detectable BH-LC binaries increases with $P_{\mathrm{max}}$ increasing, the BH and LC mass distributions have similar shapes for different $P_{\mathrm{max}}$. In summary, the choices of $P_{\mathrm{min}}$ and $P_{\mathrm{max}}$ do not much affect the BH and LC mass distributions of detectable BH-LC binaries.

\section{Discussion}
\label{sec:discussion}
\subsection{Comparisons with Previous Studies}
Here, we compare the results with the previous work.
Though \cite{Breivik2017} did not include some of the constraints we
employed, we obtained similar results to those \cite{Breivik2017}
obtained for the rapid model: FB kick does not affect both the
detectability and the parameter distributions of detectable binaries
very much.
\cite{Shao2019} considered only interstellar extinction.
Since we adopted two more observational constraints as well as the extinction,
it is reasonable that our detectability is significantly smaller than they estimated.

\cite{Yamaguchi2018} and \cite{Yalinewich2018}, which
used the same constraints as ours, estimated more than 10 times larger
number of BH-LC binaries can be detected.  They also reported
different LC mass distributions from ours.
\cite{Yamaguchi2018} indicated LC mass distributions are biased to 
heavier values such as $\sim 10$ -- $20 \msun$.
\cite{Yalinewich2018} also suggested a double-peak distribution around $\sim 1 \msun$ 
and $\sim 30 \msun$.
It is difficult to specify reasons for the difference between their and our results. They adopted analytic models, while we performed binary population synthesis calculations. There are many different points, such as single star evolution models, stellar winds, CE parameters, and tidal evolution models.

\cite{Chawlaetal2021} employed both the rapid and the delayed models with including FB kick, similarly to us. They estimated the number of detectable BH-LC binaries larger than we estimated by 3--10 times. We conclude that our results would be similar to theirs, despite of different observational constraints adopted.
They expected that \textit{Gaia} observations could give a constraint on binary evolution models, such as SN models, which is consistent with our results. Moreover, our systematic study for binary evolution models shows that the CE efficiency $\alpha$ could be also constrained, and that the natal kick model could be identified if the delayed model is correct.

\subsection{Model Uncertainties}
In the eccentricity distribution, we found that for most of the parameter sets, the detectable BH distributions are biased to larger values.
Here we discuss the above argument that eccentric and long-period BH-LC binaries tend to preserve their initial eccentricities, considering an uncertainty of tidal evolution in binaries. \cite{Yoonetal2010}\citep[see also ][]{Qinetal2018} suggested another formula for dynamical tide, and \cite{Kinugawa2020} and \cite{Tanikawa2022} took into account it in their binary population synthesis to investigate formation of merging binary BHs. \cite{Tanikawa2022} found that their dynamical tide is less efficient than that in the original BSE for MS and post-MS stars, and more efficient for He stars.
Since the majority of detectable BH-LC binaries have MS stars as LCs, eccentricities of long-period BH-LC binaries should be well preserved even if we employ the formula in \cite{Yoonetal2010}. Thus, the above argument would not be sensitive to at least an uncertainty of tidal evolution in binaries.

In this work, we only consider the solar metallicity. 
If we include lower metallicity than the solar value (hereafter, the lower metallicity case), stars will not lose their masses compared to the solar metallicity case and lighter primary stars can form BHs. 
Thus, more BH-LC binaries will be formed and the detectability will increase.
For BH mass distribution, more massive BHs such as $m_\mathrm{BH} \geq 20 \msun$ can be formed with both SN models, but the lower mass gap BHs can be formed only with the delayed model.  
Based on the important statement that the detectable BH mass distribution is similar to the Galactic one and the distribution obtained from SSE, \textit{Gaia} will detect the lower mass gap BHs if SN model is consistent with the delayed model.
The strength of FB kick should be weakened in the lower metallicity case and the effect of the kick might be somehow blurred. Although the dependence of BH-LC binary parameters on the CE efficiency $\alpha$ might be different between the solar and lower metallicity cases, the detailed discussion is beyond the scope of this work.

\section{Conclusion}
\label{sec:concl}
We simulated BH-LC binaries with a binary population synthesis code {\sf
BSE} and investigated how the detectability of BH-LC binaries with
{\it Gaia} is affected by binary parameters such as SN models, CE efficiency, and natal kick.  We employed two SN models, the rapid and
the delayed models.  The intrinsic BH mass gap is produced in the
former model, and not in the latter model.  We also adopted
three different CE efficiency $\alpha = 0.1, 1,$ and $10$ and checked
the effect of natal kick by switching it on and off.

We estimated \defbone~-- \denoten~BH-LC binaries can be observed in the five-year
observation with {\it Gaia}.  In each model, we found that parameter
distributions of the detectable binaries are roughly consistent with
the Galactic distributions. 
There are three important implications from our results:
\begin{enumerate}
\item from the BH mass distribution, if the lower mass gap is not intrinsic, {\it{Gaia}} will observe $\leq 5 \msun$ BHs. 
\item when we observe BH-LC binaries with short orbital periods and light LCs, a large CE efficiency should be favored.
\item while the rapid model is robust to FB kick, the delayed model is greatly affected by the kick. We may be able to distinguish the existence of the kick from the eccentricity distribution of detectable BH-LC binaries.
\end{enumerate}

We also investigated whether choices of $P_\mathrm{min}$ and
$P_\mathrm{max}$ will affect the detectability and the parameter
distributions.  With respect to the detectability and the parameter
distributions, we found that the choice of $P_\mathrm{min}$ (\textit{i.e.} the cadence of the astrometric observation) does not matter.  By contrast, the choice
of $P_\mathrm{max}$ can affect the detectability drastically.  If we
extend the maximum period up to 20~years from 10~years, we will find
more than three times larger number of BH-LC binaries.
If the {\it Gaia} operation is extended long enough and/or the method to identify BH-LC binaries from the {\it Gaia} data is more sophisticated, we can expect that we may find BH-LC binaries with orbital periods longer than 10~years.

\section*{Acknowledgement}
N.K. acknowledges support by the Hakubi project at Kyoto University. 
This research was supported in part by Grants-in-Aid for Scientific Research (17H06360, 19K03907) from the Japan Society for the Promotion of Science, and MEXT as “Program for Promoting Researches on the Supercomputer Fugaku” (towards a unified view of the universe: from large scale structures to planets, revealing the formation history of the universe with large-scale simulations and astronomical big data).


\bibliography{reference}{}
\bibliographystyle{aasjournal}



\end{document}